\documentclass[a4paper,11pt]{article}

\pdfoutput=1 

\usepackage[T1]{fontenc} 

\usepackage[]{caption}
\usepackage{makeidx}
\usepackage{multirow}
\usepackage{multicol}
\usepackage[dvipsnames,svgnames,table]{xcolor}
\usepackage{graphics}
\usepackage{epstopdf}
\usepackage{ulem}
\usepackage{hyperref}
\usepackage{amsmath}
\usepackage[compat=1.0.0]{tikz-feynman}
\usepackage{amssymb}
\usepackage{float}
\usepackage{fancyhdr} 
\usepackage{enumitem}
\usepackage{tcolorbox}
\usepackage{lipsum}
\usepackage{bold-extra}
\usepackage{wrapfig}
\usepackage{caption}
\usepackage[numbers]{natbib}
\allowdisplaybreaks
\usepackage{upgreek}	

\usepackage{bm}					
\usepackage{booktabs} 

\newcommand{\be}{\begin{eqnarray}}
\newcommand{\ee}{\end{eqnarray}}
\newcommand{\bea}{\begin{eqnarray}}
\newcommand{\eea}{\end{eqnarray}}

\newcommand{\ba}{\begin{array}}
\newcommand{\ea}{\end{array}}
\newcommand{\bd}{\begin{displaymath}}
\newcommand{\ed}{\end{displaymath}}
\newcommand{\beq}{\begin{equation}}
\newcommand{\eeq}{\end{equation}}

\newcommand{\nn}{\nonumber}

\def\q2 {q^2}

\def\bt{\begin{table}}
\def\et{\end{table}}

\definecolor{shilamagenta}{rgb}{0.8, 0.0, 0.8}

\definecolor{shilagreen}{rgb}{0.0, 0.5, 0.0}
\definecolor{shilacyan}{rgb}{0.0, 0.58, 0.71}

\definecolor{midnightblue}{rgb}{0.1, 0.1, 0.44}

\usepackage{hyperref}

\hypersetup{citecolor=blue,        
    linkcolor=red,  
    filecolor=magenta,      
    urlcolor=shilacyan  
}

\usepackage{cleveref}
\usepackage{slashed}

\begin{document}

\title{\textcolor{red}{Study of entropy production due to electroweak phase transition in $Z_2$ symmetric extension of the Standard Model}}

\author{Arnab~Chaudhuri
$^{a,b}$\footnote{{\bf e-mail}: \href{mailto:arnabchaudhuri.7@gmail.com}{arnabchaudhuri.7@gmail.com} and \href{mailto:arnab.c@iopb.res.in}{arnab.c@iopb.res.in}},
and
Jaydeb Das$^{c}$\footnote{{\bf e-mail}: \href{mailto:jaydebphysics@gmail.com}{jaydebphysics@gmail.com}}
\\
$^a$ \small{Institute of Physics, Bhubaneswar,} 
\small{Bhubaneswar 751005, India.}\\
$^b$ \small{Homi Bhabha National Institute,} 
\small{ Mumbai 400085, India.}\\
$^c$ \small{Department of Physics and Astrophysics, University of Delhi,}\\\small{Delhi 110007, India.}\\
}

\date{}
\maketitle

\begin{abstract}
In this work we consider the simple $Z_2$ symmetric extension to the Standard Model (SM) and proceed to study the nature of electroweak phase transition (EWPT) in the early universe. We show that the nature of the phase transition changes from a smooth crossover in the SM to a strong first order with this addition of the real scalar. Furthermore, we show the entropy release in this scenario is higher than that of the SM. This can lead to a strong dilution of frozen out dark matter particles and baryon asymmetry, if something existed before the onset of the phase transition. 
\end{abstract}

\section{Introduction}
The Standard Model (SM) of particle physics is successful in describing three of the four known fundamental forces (electromagnetic, weak and strong) in the universe and classifying all known elementary particles. Although the SM is one of the most popular theories present at the current moment, it leaves some phenomena unexplained. It falls short of being a complete theory of fundamental interactions. For example, it does not fully explain baryon asymmetry, incorporate the full theory of gravitation, \cite{SC} as described by general relativity, or account for the universe's accelerating expansion as possibly described by dark energy and further does not contain any viable candidate for dark matter. 

The predominance of matter over antimatter, i.e., the mechanism of  baryogenesis follows Sakharov's principle \cite{Sakh} and they are \textbf{i) }Non-conservation of Baryon numbers, \textbf{ii) }Breaking of C and CP invariance, and \textbf{iii) }Deviation from thermal equilibrium. For a successful explanation of the baryon asymmetry \cite{A,B} in the universe through baryogenesis, a strong first-order electroweak phase transition (EWPT) in the early universe is necessary. Cosmic EWPT happened when the hot universe cooled down enough in the primeval time so that the potential of the Higgs field settled at a non-zero minimum and in consequence, the symmetry of the theory $SU(3)_C \times$ $SU(2)_L\times U(1)_Y$  broke to $U(1)_{\rm em}$. At the time of first-order EWPT, bubbles of the broken phase originate and baryon-antibaryon asymmetry generates outside the wall of the bubbles of the broken phase. However after the discovery of the SM  Higgs boson \cite{Higgs}, it became more obvious that EWPT within the framework of the SM is a  smooth crossover phase transition, see \cite{Chaudhuri:2017icn, shapas}. Hence for a successful electroweak baryogenesis (EWBG), the theory of EWPT should be of first order and hence a theory beyond the SM is necessary. Strong first order phase transition in $Z_2$ and $Z_3$ extension of the SM are studied in \cite{Morrissey:2012db}, \cite{z31}, \cite{z32} and \cite{z33}.

On the other hand, $\sim 26.5\%$ of the total energy density of the universe is contributed by the dark matter (DM) whose origin is a mystery till date. Even though, primordial black holes (PBH) and MACHOs which are considered as one of the viable baryonic DM candidates, it is now clear that they are unable to contribute completely to the 
DM energy density of the universe. There are theories about multicharged extension of the Standard Model like dark atoms which can be viable dark matter candidates, ~\cite{Chaudhuri:2021ppr}. But there are no experimental evidences as of now. Not only the SM cannot explain the phenomenon of successful baryogenesis, but there are no irrefutable theories in SM about nonbaryonic DM particle which can successfully explain all the observations. 

Due to these short comings of the SM, the search for beyond Standard Model (BSM) physics has become a hot topic among the physicists now a days. For them the recent result from Fermilab about $g_\mu -2$ for muon may be a ray of hope. $g_\mu$ is the gyromagnetic ratio of muon which is defined as the ratio of magnetic moment to the angular moment of muon and whose value is $2$ from tree-level calculation. If we define $a_\mu = (g_\mu-2)/2$, then higher order loop corrections from SM gives $a_\mu = 116,591,810(43)~\times~10^{-11}$ where the value measured from Fermilab is $16,592,061(41)~\times~10^{-11}$ which differs from SM at $3.3\sigma$ level~\cite{Muong-2:2021ojo}. This contradiction is actually buttressed the previously claimed result from the E821 experiment at Brookhaven National Lab (BNL). There are numerous explanation for this anomalous result including the existence of BSM.

The early universe is often associated with thermal equilibrium and negligible chemical potential. During the course of universe expansion, the entropy density per comoving volume was conserved and the conservation law follows:
\be \label{entropy}
s=\frac{\rho+\mathcal{P}}{T}a^3=\rm const,
\ee
where $\rho$ and $\mathcal{P}$ are the energy and the pressure density of the plasma respectively, $T$ is the temperature of the plasma and $a$ is the cosmological scale factor. In thermal equilibrium the distribution function of any particle species is determined by the chemical potential $\mu_i$ of each particle types and the temperature of the plasma. But in the case of thermal in-equilibrium the entropy conservation law breaks down. Several instances of entropy non-conservation includes QCD phase transition at $T\sim 150$~MeV, for review \cite{QCD}. If PBHs of sufficiently small masses dominated the universe at a certain point during the course of  universe evolution, the evaporation of such PBHs can lead to the influx of sufficient entropy into the plasma. For details and review see \cite{ACAD} and \cite{ADPDIN}.

Another source of entropy production, which is the primary motive of this paper, is due to EWPT. Possibly the largest entropy release in the Standard Model took place in the process of
the EWPT from symmetric to asymmetric electroweak phase in the course of the cosmological cooling down. In principle, the transition could be either first order or second order, even very smooth crossover. Within the framework of the SM with single Higgs the phase transition is crossover in nature. But in extended theories of the SM, it can be of first-order, \cite{simone}, \cite{HN}, \cite{BKS}, \cite{ACMK}, \cite{ACMKSP1} and \cite{ACMKSP2}. 

In this work we consider a $Z_2$ symmetric singlet scalar extension to the SM. We show that this simple inclusion changes the nature of the EWPT from a smooth crossover to a first-order phase transition. We further proceed to calculate the entropy release during the first order phase transition and show that it is considerably higher than the SM scenario even with a single singlet scalar extension. The paper is arranged as follows: In the second section the details about the potential along with the correction terms are shown. The next section section gives details about the nature of the phase transition and the calculation of entropy production. This is followed by a generic conclusion and discussion.

\section{Tree-level Potential}
\label{sec:tree}
In our framework, we extend the SM by a real scalar which transforms as a singlet under the SM gauge group. We call this extra singlet bosonic degree of freedom as $S$.  In addition to the SM gauge symmetry, we impose an extra $Z_2$ symmetry i.e $S\to -S$ to the scalar potential, so that we can exclude all odd power terms of $S$ in the Lagrangian. So tree level potential consists of pure Higgs potential of the SM along with  quadratic and  quartic term of $S$ and a portal term which is essentially an interaction term between singlet scalar and SM Higgs field. So the renormalizable tree level potential with a real singlet scalar extension to the Standard Model consists of scalar field $S$ and Higgs doublet $\phi$ is given by:\\
\begin{equation} \label{tree}
\begin{split}
V (\phi, S) = - \mu_h^2 \phi^{\dagger} \phi + \lambda_h ( \phi^{\dagger} \phi )^2 - \frac{1}{2} \mu_S^2 S^2 + \frac{1}{4} \lambda_S S^4 + \frac{1}{2} \lambda_m S^2 (\phi^{\dagger} \phi) ,
\end{split}
\end{equation}
In the above equation \ref{tree}, $\mu_h^2 $ and $ \mu_S^2$  are the bare parameters with mass dimension 2, $\lambda_h$ and $\lambda_S$ are the dimensionless quartic coupling constants for Higgs doublet $\phi$ and singlet scalar field $S$ respectively. There is another dimensionless coupling constant $\lambda_m$, called Higgs portal coupling related to SM Higgs and singlet scalar interaction term in the tree level potential. Because of the extra $Z_2$ symmetry, we do not include the terms which are linear and cubic in $S$ in the potential. 

The SM Higgs doublet $\phi$ can be written as:
\begin{equation} \label{doub}
\begin{split}
\phi = \frac{1}{\sqrt{2}} \begin{pmatrix} \chi_1+i\chi_2 \\ h+i\chi_3 \end{pmatrix},
\end{split}
\end{equation}
where $\chi_1$, $\chi_2$, $\chi_3$ are three Goldstone bosons, and $h$ is the Higgs boson. The tree-level potential in terms of the classical background fields $h$ and $S$ reads as:
\begin{equation} \label{4}
\begin{split}
V_0 (h, S) = -\frac{1}{2} \mu_h^2 h^2 + \frac{1}{4} \lambda_h h^4 - \frac{1}{2} \mu_S^2 S^2 + \frac{1}{4} \lambda_S S^4 + \frac{1}{4} \lambda_m S^2 h^2.
\end{split}
\end{equation}
 It is needed to be mentioned here that the classical background fields $h$ and $S$ in equation \ref{4} are not same as that of equation \ref{tree} and \ref{doub}. After expanding the potential around the classical background fields we obtain the above equation \ref{4}.

We are interested in studying the strong first-order electroweak phase transition in that scenario where at higher temperature electroweak symmetry is there but discrete $Z_2$ symmetry is spontaneously broken by VEV of singlet scalar. When temperature gradually decreases SM Higgs field gets VEV but singlet scalar has no VEV. In other word, at lower temperature, $Z_2$ symmetry is there but SM gauge symmetry is spontaneously broken by VEV of Higgs field. Therefore, at zero temperature SM Higgs has VEV, $v_{\rm EW} = 246$ GeV but singlet scalar has no VEV. So all bare parameters can be written in terms of physical measurable quantities at zero temperature:
\begin{eqnarray}
\mu_h^2 = \frac{m_h^2}{2},\,\ \lambda_h = \frac{\mu_h^2}{v_{\rm EW}^2},\,\,\ \mu_s^2 = - m_S^2 + \frac{1}{2}\lambda_m v_{\rm EW}^2
\end{eqnarray}
So the input parameters for this scenario are mass of the Higgs boson, $m_h = $125 GeV, the VEV of Higgs field, $v_{\rm EW}=$ 246 GeV, mass and quartic coupling of the singlet scalar, $m_S$ and   $\lambda_S$ respectively and Higgs portal coupling, $\lambda_m$. Here we consider the zero temperature one-loop Coleman-Weinberg (CW) potential in the dimensional regularisation schemes to avoid the infrared divergences which appear from the massless Goldstone modes at zero temperature in the on-shell renormalization schemes. For detailed calculations on infrared divergence, see \cite{resum}.
We will talk about zero temperature CW potential in both on-shell as well as dimensional regularisation schemes in latter section.  
\subsection{Field dependent masses}
In general way, 2$\times$2 symmetric mass-squared matrix for Higgs and singlet scalar in terms of classical background field can be written as:
\begin{equation} \label{massT}
\begin{split}
M^2 (h,S)  &= \begin{pmatrix} \frac{\partial^2 V}{\partial h^2} &\frac{\partial^2 V}{\partial h \partial S} \\  \frac{\partial^2 V}{\partial h \partial S} & \frac{\partial^2 V}{\partial S^2}  \end{pmatrix}  \equiv \begin{pmatrix} M_{hh}^2(h,S) & M_{hS}^2(h,S) \\  M_{hS}^2(h,S) & M_{SS}^2(h,S) \end{pmatrix}  .
\end{split}
\end{equation}
where 
\begin{equation}
M_{hh}^2 (h,S)= 3 \lambda_h h^2 - \mu_h^2+ \frac{1}{2} \lambda_m S^2 ,\,\ M_{SS}^2 (h,S) = \frac{1}{2} \lambda_m h^2  - \mu_S^2 + 3 \lambda_S S^2,\,\ M_{hS}^2 (h,S) = \lambda_m hS.
\end{equation}
After diagolization of the above mass-squared matrix, $M^2(h,S)$, the eigenvalues of the mass matrix give the field dependent masses of Higgs and singlet scalar:     
\begin{eqnarray}\label{eq:mass1}
 m_{h_1}^2 (h,S) = \frac{1}{2}\Big\{ m_{hh}^2 (h,S) + m_{SS}^2 (h,S) + \sqrt{\big(m_{hh}^2 (h,S) - m_{SS}^2 (h,S)\big)^2 - 4 m_{hS}^4 (h,S)}\Big\} \nn\ \\
  m_{h_2}^2 (h,S) = \frac{1}{2}\Big\{ m_{hh}^2 (h,S) + m_{SS}^2 (h,S) - \sqrt{\big(m_{hh}^2 (h,S) - m_{SS}^2 (h,S)\big)^2 - 4 m_{hS}^4 (h,S)}\Big\}
\end{eqnarray}
The field dependent masses of other degrees of freedom are given by:
\begin{equation}\label{eq:mass2}
\begin{split}
&m_W^2 (h,S) = \frac{g^2}{4} h^2,\quad m_Z^2 (h,S)= \frac{g^{'2}+g^2}{4} h^2,\quad m_t^2 (h,S) = \frac{1}{2} y_t^2 h^2,\\
&m_{\chi_{1,2,3}}^2 (h,S) =-\mu_h^2+ \lambda_h h^2 + \frac{1}{2} \lambda_m S^2,\\
\end{split}
\end{equation}
where $y_t$ is the Yukawa coupling for top quark,  $\chi_{1,2,3} $  are the Goldstone bosons as mentioned above. $m_W(h,S)$, $m_Z(h,S)$ and $m_t(h,S)$ are the field dependent masses of W boson, Z boson and top quark respectively.


From perturbative unitarity condition, the coupling constants are constrained in the following ways:
\begin{eqnarray}
\lambda_h < 4\pi,\,\,\ \lambda_S < 4\pi,\,\,\,\ |\lambda_m | < 8\pi,\,\,\  3\lambda_h + 2\lambda_S + \sqrt{(3\lambda_h - 2\lambda_S)^2 + 2\lambda_m^2} < 8\pi
\end{eqnarray}
For the potential to be bounded from below, the vacuum stability conditions should follow:
\begin{eqnarray}
\lambda_h >0,\,\,\ \lambda_S >0,\,\,\ \lambda_m  > -2\sqrt{\lambda_h \lambda_S}
\end{eqnarray}
Detailed calculations for perturbative unitarity can be found in \cite{18,19} 
and for vacuum stability in \cite{VS}.

\section{One-loop effective potential at finite temperature}
The one-loop effective potential at non zero temperature is given by \cite{eff}:
\begin{equation} \label{eq:finiteT}
\begin{split}
V^T_{\rm 1-loop} (h, S, T)
&= \frac{T^4}{2\pi^2} \left[ \sum_{B} n_B J_B \left(\frac{m_B^2(h,S)} {T^2}\right) + \sum_{F} n_F J_F \left(\frac{m_F^2(h,S)} {T^2}\right)  \right],
\end{split}
\end{equation}
Here, B stands for all bosonic degrees of freedom that couple directly to Higgs boson, therefore, B = $\{W, Z, h_1, h_2, \chi_{1,2,3}\}$ and F stands for only top quark fermion. $J_B$ and $J_F$ are the thermal bosonic and fermionic functions which are given as follow.
\begin{eqnarray}
J_{B} \left(\frac{m_B^2(h,S)} {T^2}\right)& = & \int_0^\infty dx x^2 \log\left( 1 - e^{-\sqrt{x^2+\frac{m_B^2(h,S)} {T^2}}}\right) \\
J_{F} \left(\frac{m_F^2(h,S)} {T^2}\right)& = & \int_0^\infty dx x^2 \log\left( 1 + e^{-\sqrt{x^2+\frac{m_F^2(h,S)} {T^2}}}\right)
\end{eqnarray}
We consider only top quark contribution here since other fermionic contribution of the SM are less dominant because of small Yukawa coupling. If field dependent masses at a given temperature are very less than temperature of the plasma i.e $\frac{m_i^2}{T^2}<< 1$, thermal function admits a high-temperature expansion which will be very useful for practical applications. It is given by \cite{eff}:

\begin{eqnarray} \label{eq:expn1}
&& J_B \left[\frac{m_B^2}{T^2} \right] = -\frac{\pi^4}{45} + \frac{\pi^2}{12T^2} m_B^2 - \frac{\pi}{6T^3} (m_B^2)^{3/2} - \frac{1}{32T^4}m_B^4 \ln \frac{m_B^2}{a_b T^2} + \cdots,   \\ \label{eq:expn2}
&&J_F \left[\frac{m_F^2}{T^2} \right] =  \frac{7\pi^4 }{360} - \frac{\pi^2}{24T^2} m_F^2 - \frac{1}{32T^4}m_F^4 \ln \frac{m_F^2}{a_f T^2} + \cdots, 
\end{eqnarray}
Where $a_f = \pi^2\exp \left(3/2-2\gamma_E\right)$ and $a_b = 16 \pi^2\exp \left(3/2-2\gamma_E\right)$, Euler constant, $\gamma_E = 0.577$. In equation \ref{eq:finiteT}, $n_B$ and $n_F$ represent the degrees of freedom of bosons and fermions,
\begin{eqnarray}\label{eq:degrees}
n_W = 6,\,\,\ n_Z = 3,\,\,\ n_{h_1} = 1,\,\ n_{h_2} = 1,\,\,\ n_{\chi_{1,2,3}}=1,\,\,\ n_t = -12.
\end{eqnarray}
For another limiting case, at a given temperature of plasma, if the field dependent mass of a particle is much higher than temperature i.e $\frac{m_i(h,S)^2}{T^2}>>1$, then the thermal functions, both bosonic and fermionic, behave like exponentially decreasing function \cite{22}. Therefore,  in thermal effective potential, the contribution of a particle with field dependent mass greater than that of the temperature is almost negligible. The first terms on the right hand side of equations \ref{eq:expn1} and \ref{eq:expn2} are independent of classical background fields, therefore, these terms are irrelevant in calculating critical temperature, $T_c$.  

In dimensional regularization scheme, the temperature-independent Coleman-Weinberg (CW) potential term of the effective potential at one-loop order is given by \cite{eff}:
\bea
V_{\rm 1-loop}^{\rm CW} (h,S)
&= \frac{1}{64\pi^2} \left(  \sum_{B}n_B m_B^4(h,S) \Big[ \log \Big( \frac{m_B^2(h,S)}{Q^2} \Big) - c_B \Big]\right. \nonumber \\
&\left. +  \sum_{F}  n_F m_F^4(h, S) \Big[ \log \Big( \frac{m_F^2(h,S)}{Q^2} \Big) - \frac{3}{2} \Big] \right),
\eea
where $B$ and  $F$ have been defined above and  $c_B = 3/2$ and $5/6$ for scalar and vector bosons respectively. $Q$ is the renormalization scale of the theory which is taken to be $v_{\rm EW}=246$ GeV, the vev of the Higgs field at zero temperature. Field dependent masses of particles, $m_i(h,S) = \{m_B(h,S), m_F(h,S)\}$, are given in \ref{eq:mass1}, \ref{eq:mass2} and all degrees of freedom, $n_i = \{n_B, n_F\}$, are given in \ref{eq:degrees}.

A very useful scheme, called cut-off regularization scheme, is obtained by regularizing the theory with a cut-off. Here we add counter terms in the potential in such a way such that the minima of the Higgs potential at $v_{\rm EW} = 246$ GeV, remains unchanged and Higgs and singlet scalar masses remain unchanged with respect to tree level potential. So the CW potential in cut-off regularization scheme reads \cite{eff},

\begin{equation} \label{}
\begin{split}
V_{\rm 1-loop}^{\rm CW} (h,S)
= \frac{1}{64\pi^2} \sum_{i=\{B\},\{F\}} n_i \Big\{ & m_i^4(h,S) \Big[ \log \frac{m_i^2(h,S)}{m_i^2(v_{\rm EW},0)}  - \frac{3}{2} \Big] 
 + 2 m_i^2(h,S)m_i^2(v_{\rm EW},0)  \Big\},
\end{split}
\end{equation}
where $m_i^2(v_{\rm EW},0)$ are the square of the masses of particles at electroweak vacuum i.e at $v_{\rm EW} =$ 246 GeV and degrees of freedom, $n_i$ and field dependent mass-squared, $m_i^2(h,S)$ are given in \ref{eq:degrees} and \ref{eq:mass1}, \ref{eq:mass2} respectively.
\subsection{Thermal resummation}
In order to reassure the validity of the one-loop perturbative expansion, the corrections from the daisy re-summation should be included in the one-loop potential. The leading order resummation results give thermal corrections of $\Pi_i = d_i T^2$ to effective masses where $d_i$ for different degrees of freedom in the plasma are are given in eq. \ref{di} . So in our later numerical analysis we have replaced bosonic masses, $m_i^2(h, S) \to m_i^2(h,S, T) =  m_i^2(h, S) + d_i T^2$, where $d_iT^2$ is the finite temperature contribution to the self-energies \cite{23}.
\begin{equation} \label{di}
\begin{split}
&d_{\chi_i} ={ \frac{3}{16} g^2 + \frac{1}{16} g^{'2}  } + \frac{1}{2} \lambda_h + \frac{1}{4} y_t^2 + \frac{1}{24} \lambda_m,\\
&d_{hh} ={ \frac{3}{16} g^2 + \frac{1}{16} g^{'2}  } + { \frac{1}{2} } \lambda_h + \frac{1}{4} y_t^2 + \frac{1}{24} \lambda_m,\quad 
d_{SS} =\frac{1}{4} \lambda_S + \frac{1}{6} \lambda_m,\quad 
d_{Sh} \approx 0.
\end{split}
\end{equation}
Other than the self-energies correction to masses of Higgs boson, singlet scalar and Goldstone bosons, the thermal correction to masses of electroweak gauge bosons have been discussed in \cite{24,25}. As the temperature correction to the masses of electrweak gauge bosons is very small compared to the field dependent but temperature independent mass terms, therefore we do not consider the thermal correction of the electroweak gauge bosons in our numerical analysis. On the other hand, fermions can not receive thermal correction to the masses because of gauge symmetry.
A general and more rigorous treatment of the thermal re-summation in calculating $\Pi_i$ in the one-loop effective potential at finite temperature can be found in \cite{26}.
 
Here we want to discuss the effective potential without one-loop CW term and thermal re-summation contributions at a very high temperature for analytic calculation purposes. At very high temperatures where $m_i^2(h, S)/T^2 << 1$, we can take the high-temperature expansion of thermal functions that we have discussed before. But for better analysis of phase transition, we consider the full effective potential composing tree-level term, one-loop CW term, and one-loop temperature corrected term without taking any high-temperature approximation. Therefore, without considering the complicated CW term and daisy re-summation contributions, in terms of fields dependent part of one-loop effective potential at very high-temperature becomes:

\begin{equation} \label{eq:hightem}
\begin{split}
V(h,S, T) &= V_0 (h,S) + V_{\rm 1-loop}^{T} (h,S, T)  \\
&\approx -\frac{1}{2}\mu_h^2(T)h^2 + \frac{1}{4}\lambda_h h^4 - E^{\rm SM} T h^3 -  \frac{1}{2} \mu_S^2(T) S^2 +  \frac{1}{4} \lambda_S S^4 +  \frac{1}{4} \lambda_m h^2 S^2- E(h,S) T,
\end{split}
\end{equation}
where the mass-squared parameters are defined by
\begin{equation} \label{eq:hightem1}
\begin{split}
\mu_h^2(T) = \mu_h^2 - c_h T^2,\,\,\ \mu_S^2(T) = \mu_S^2 - c_S T^2,
\end{split}
\end{equation}
with $\mu_h^2$ and $\mu_S^2$ the mass-squared defined at $T=0$.\\
All coefficients in the above equation \ref{eq:hightem} and \ref{eq:hightem1} are given by
\begin{equation} \label{eq:htvcoe}
\begin{split}
&  c_h = \frac{1}{48} [9 g^2 + 3 g^{'2} + 2 (6 y_t^2 + 12 \lambda_h +\lambda_m)],  \\
&E^{\rm SM} =  \frac{1}{32\pi} \Big[ 2 g^3 +  \sqrt{g^2+g^{'2}}^3 \Big] ,\\
&  c_S =  \frac{1}{12} (2\lambda_m + 3 \lambda_S  ) ,  \\
& E(h,S) =  \frac{1}{12\pi} \Big[   \big(m_{h_1}^2(h,S)\big)^{3/2} +\big(m_{h_2}^2(h,S) \big)^{3/2} + 3\big(m_{\chi_1}^2(h,S)\big)^{3/2}  \Big],
\end{split}
\end{equation}
where $m^2_{h_{\rm 1,2}}(h,S)$ and $m_{\chi_1}^2(h,S)$ are given in equations ~(\ref{eq:mass1}) and \ref{eq:mass2}.

\begin{figure}
    \centering
    \includegraphics[scale=0.41]{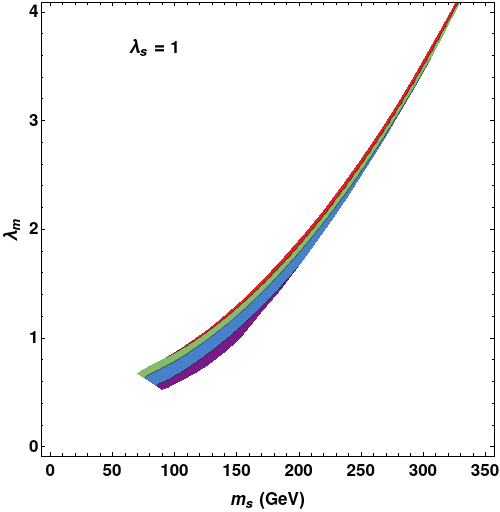}
    \includegraphics[scale=0.42]{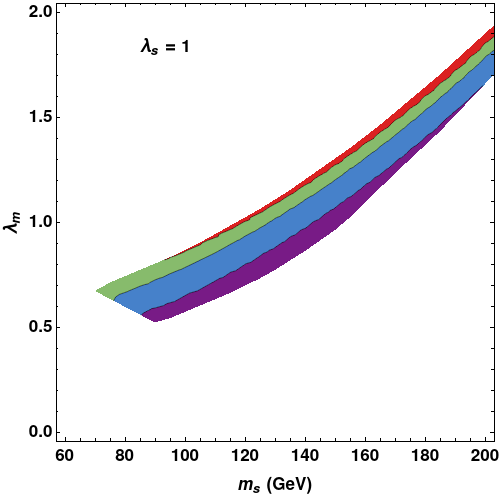}
    \includegraphics[scale=0.42]{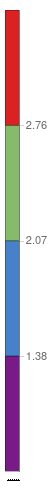}
    \caption{Allowed region in $\lambda_m $ and $m_S$ (GeV) plane for $\lambda_S$ = 1 where different color represents the phase transition strength. The right panel shows the zoomed version of the left panel upto $\lambda_m=2$. \label{fig:lams0.1}}
\end{figure}
\begin{figure}
    \centering
    \includegraphics[scale=0.4]{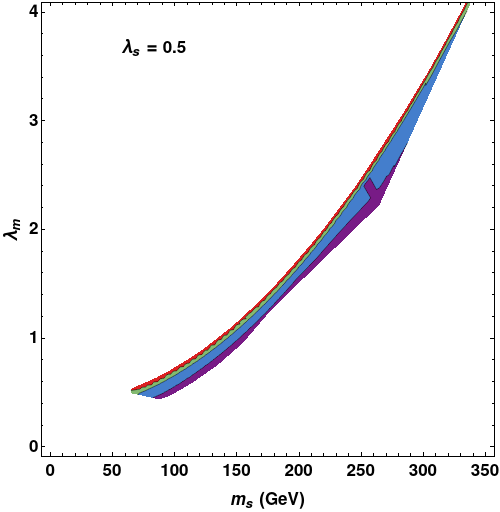}
    \includegraphics[scale=0.4]{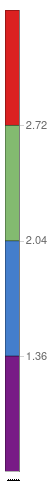}
    \includegraphics[scale=0.4]{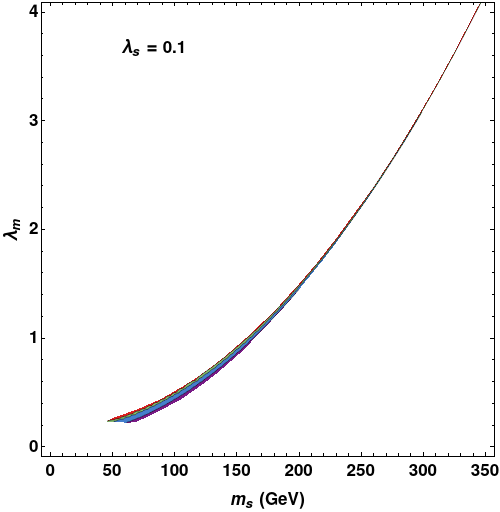}
    \includegraphics[scale=0.4]{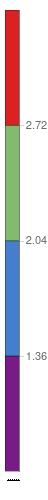}
    \caption{Allowed region in $\lambda_m $-$m_S$ (GeV) plane for $\lambda_S$ = 0.5 (left panel) and $\lambda_S$ = 0.1 (right panel) provided first-order EWPT. Here different color represents the phase transition strength, $v_c/T_c$.  \label{fig:lams1}}
\end{figure}

\section{Numerical Analysis}
The total effective potential can be written as 
\begin{eqnarray}\label{eq:effpot}
V_{\rm eff}^{\rm tot}(h,S,T) = V_0(h,S) + V_{\rm 1-loop}^{\rm CW}(h,S,T) + V_{\rm 1-loop}^T(h,S,T)
\end{eqnarray}
It is needed to be mentioned that $ V_{\rm 1-loop}^{\rm CW}(h,S,T$)  depends on the temperature explicitly due to the thermal re-summation of masses from the daisy diagrams to maintain the  validity of the perturbative expansion at one-loop level. Effectively all three terms on the right hand-side of the equation \ref{eq:effpot} also depend on the temperature implicitly through the VEV of Higgs, $v$ and the singlet scalar, $w$. 

Here we are interested in studying the first-order EWPT in that scenario where the phase transition occurs through three steps processes. At very high temperature both the Higgs and the singlet scalar fields have no VEV, i.e, universe was in fully symmetric phase. When the temperature gradually decreases, the  singlet scalar field $S$ gets VEV, $w$ but Higgs field does not get VEV, i.e $Z_2$ symmetry breaks spontaneously but electroweak symmetry is there. When temperature further decreases, the Higgs field gets VEV, $v$ but singlet scalar has no VEV i.e, electroweak symmetry breaks spontaneously but $Z_2$ symmetry restores. For a given parameter space, the conditions for calculating critical temperature, $T_c$ and VEV of Higgs field, $v_c$ at $T_c$ are given by
\begin{eqnarray}\label{eq:effsol}
V_{\rm eff}^{\rm tot}(v_c,0,T_c)& = & V_{\rm eff}^{\rm tot}(0,w_c,T_c)\\ \label{eq:dervsol}
\frac{dV_{\rm eff}^{\rm tot}(h,S,T_c)}{dh}\big|_{h=v_c, S=0,T_c} &=& 0,\,\ \frac{dV_{\rm eff}^{\rm tot}(h,S,T_c)}{ds}\big|_{h=0, S=w_c,T_c} =0.
\end{eqnarray}
Here $v_c$ and $w_c$ are the VEV of Higgs field and singlet scalar field respectively at the critical temperature. Analytically it is impossible to solve the above three equations \ref{eq:effsol} and \ref{eq:dervsol} simultaneously. So we do a scan over the parameter space numerically imposing the condition \ref{eq:1stcon} for first-order EWPT. Necessary condition for first order EWPT is that
\begin{eqnarray} \label{eq:1stcon}
\frac{v_c}{T_c}\geq 1.
\end{eqnarray}
In order to calculate the $T_c$ and $v_c$ from eqs. \ref{eq:effsol} and \ref{eq:dervsol} we cannot get the exact solution of these three above mentioned equations. Therefore, for numerical scanning, we solve the equation \ref{eq:effsol} upto $1\%$ error.
 
In figures \ref{fig:lams0.1} and  \ref{fig:lams1}, we have shown the allowed region in $\lambda_m$ and  $m_S$ parameter space  for a fixed $\lambda_S$ provided first order EWPT. The different colors of the figures represent the strength, $\frac{v_c}{T_c}$, of the phase transition. In other words, higher the value of $\frac{v_c}{T_c}$, the stronger is the phase transition.
In left panel of figure \ref{fig:lams0.1}, we have shown the variation of $\lambda_m$ with respect to $m_S$ for $\lambda_S = 1$ for first-order EWPT. From this figure we can see that first order EWPT can be achieved for higher values of $m_S$ only when the value of $\lambda_m$ is increased. For $m_S\leq 70~\rm GeV$, which corresponds to $\lambda_S=1$, we cannot observe any signature for strong first order EWPT. We can clearly understand from this figure that after a certain value of $m_S$ the parameter space gets constrained. The right panel of figure \ref{fig:lams0.1} shows a zoomed version of the left panel where a wider range in the parameter space can be found upto $\lambda_m=2$, the region almost corresponds to the electroweak scale. In this electroweak region, we get wider parameter space in $\lambda_m-m_S$ plane because if $m_S \gg$ electroweak scale then the scalar decouples from the SM Higgs. For this mass region of $m_S$, the contribution of the singlet scalar becomes negligible in the effective potential. 

In the left panel of figure \ref{fig:lams1}, we have shown the allowed parameter space in the $\lambda_m-m_S$ plane for $\lambda_S=0.5$. From this figure we can see that we cannot get any parameter space for $m_S \leq 65$ GeV. The right panel of this figure shows that the variation of $\lambda_m$ with respect to $m_S$ for $\lambda_S = 0.1$. From this figure we can see that we cannot get any allowed wider parameter space for higher values of $m_S$ with compare to $\lambda_S=1$ and $\lambda_S=0.5$ scenarios. This is because for such low value of $\lambda_S$, the potential is not bounded from below for higher values of $m_S$ and we cannot get any stable minima ($w$) for $S$. By comparing all the figures we can see an allowed parameter space in $\lambda_m-m_S$ plane for lower values of $m_S$ and $\lambda_m$ when $\lambda_S$ is also lower. Here we have considered the $\overline{{\rm MS}}$ renormalization scheme where the one-loop potential explicitly depends on the choice of renormalization scale $Q$. In order to reduce the scale-dependence, an RGE improvement should be implemented on the one-loop CW potential, \cite{32,33}. We leave this implementation for future studies.

\subsection{Calculation of entropy release into the plasma}
In order to calculate the entropy release due to EWPT, one needs to take into account the energy and pressure density of the plasma at the instant of the phase transition. The energy and the pressure density can be calculated using the energy-momentum tensor ($T_{\mu \nu}$), which is given as:
\be \label{EMT}
T_{\mu \nu}=\partial_{\mu}\phi\partial_{\nu}\phi+\partial_{\mu}S\partial_{\nu}S-g_{\alpha\beta}\left(g^{\alpha\beta}(\partial_{\alpha}\phi\partial_{\beta}\phi+\partial_{\alpha}S\partial_{\beta}S) - V_{\rm eff}^{\rm tot}(h,S,T) \right)+~\rm SM~Fermions.
\ee
The last part of equation \ref{EMT} can be found in \cite{Chaudhuri:2017icn}. The fermionic sector remains the same as that of the SM since the singlet scalar $S$ does not modify the fermonic part.

During that epoch the universe can assumed to be homogeneous and isotropic and hence the spatial derivatives of the scalar and the Higgs field can be neglected. We further assume that the scalar and the Higgs field oscillate around their minima and the damping rate of the oscillation is very high. Under these assumptions, the energy and the pressure density of the plasma is given by:
\begin{eqnarray}
\rho &=& \Dot{\phi}_{\rm min}^2+\Dot{S}_{\rm min}^2
+  V_{\rm eff}^{\rm tot}(h,S,T) + \frac{g_* \pi^2}{30} T^4.  \label{eq:energy density}  \\
\mathcal{P} &=& \Dot{\phi}_{\rm min}^2+\Dot{S}_{\rm min}^2 -  V_{\rm eff}^{\rm tot}(h,S,T) + \frac{1}{3} \frac{g_* \pi^2}{30} T^4. \label{eq:pressure density} 
\end{eqnarray}
The last terms in equation \eqref{eq:energy density} and eq.\eqref{eq:pressure density}  arise from the Yukawa interaction between fermions and Higgs bosons and from
the energy density of the fermions, the gauge bosons, and the interaction between the Higgs and gauge bosons. $g_*$ depends on the effective number of particles present in the relativistic soup at or near the EWPT. It's value in our model is greater than the value in SM.

It is mentioned before in \ref{entropy}, entropy density is conserved for relativistic species with negligible chemical potential. From eq.\eqref{eq:energy density} and equation \eqref{eq:pressure density} we get
\begin{eqnarray}
\rho + \mathcal{P} = 2 \Dot{\phi}_{\rm min}^2 +2\Dot{S}_{min}^2 + \frac{4}{3} \frac{g_* \pi^2}{30} T^4 
\end{eqnarray}
It is evident that $g_*$ will change with the decoupling process and thus $s$ for relativistic plasma will increase for our considered scenario. Then the increase in entropy can be calculated using conservation law:
\begin{eqnarray}
\dot{\rho}= - 3 H \left ( \rho + \mathcal{P} \right ) . \label{eq: solve it}
\end{eqnarray}

Using the above assumptions, equation \ref{eq: solve it} takes the approximate form:
\be
\frac{\dot{T}}{T}\left[g_*v_c^2T^2\left(1-\frac{T^2}{T_c^2} \frac{g_*(m)}{g_*} \right)+\frac{4\pi^2g_*}{30}T^4 \right]= -\frac{4H\pi^2g_*}{30}T^4,
\ee
where $T$ is the temperature of the plasma which is decreasing since the phase transition, $v_c$ is the effective value of the potential at the moment of phase transition, $g_*\approx 107.75$ is the effective degrees of freedom of the whole system, $g_*(m)$ is the reduced effective degrees of freedom when certain species became non-relativistic and $H$ is the Hubble parameter. 

Due to the presence of the new scalar $S$, analytical solution of equation \ref{eq: solve it} was not possible and we reverted to numerical solution of the differential equation. We did not take into account the modification of the evolution due to annihilation of non-relativistic species, for example, the annihilation of $e^+e^-$ which takes place below the mass of the electron. This is disregarded because if the annihilating particles are in thermal equilibrium state with vanishing
chemical potential, the entropy density in this process is conserved.

For $T\gg T_c$, the universe was in thermal equilibrium, and relativistic particles dominated the universe. The 
contribution to the overall energy
 density of the universe from those who were already massive (e.g., decoupled DM) was also insignificant.
To this extend, the entropy density per unit comoving volume follows the conservation law, as mentioned in equation \ref{entropy}.
In this scenario, the sum of the energy and the pressure density can lead up to 
\begin{equation}
\label{eq:rho+P}
\rho_r +\mathcal{P}_r \sim g_* T^4. 
\end{equation}

It is to be noted that $g_*$ is not constant but varies over time. It depends on the components of the primordial plasma. And hence,
\begin{equation}
T \sim a^{-1}.
\end{equation}

It is worthwhile to mention that the  conserved quantity is $s=g_*(T)a^3T^3$. But while estimating the amount of entropy released we calculated $a^3T^3$. This is because the contribution to entropy is dominated by the heaviest particle in the temperature range $m(h,S,T) < T$ . So for these temperatures
$g_* (T )T = \rm const.$ and the relative entropy rise is given just by $a^3 T^3$ . Since the final temperature
$T_f = m(h,S,T_f)$, below which new particle species start to dominate, is not dependent upon $g_*$ ,
the relatives increase of entropy is determined by $T^3 a^3$. Thus the rise in entropy is dominated by the change in the scale factor, i.e., $a^3T^3$, suggesting the influx of entropy happening over $g_*a^3T^3$. And hence the net entropy release is given by the generic expression: 
\be
\frac{\delta s}{s} = \frac{(a_c \, T_c)^3 - (a \, T)^3}{(a_c\, T_c)^3},
\ee 
where $a_c$ and $T_c$ are the cosmological scale factor and the temperature of the plasma when the phase transition took place.

Following the above assumptions, for two sets of benchmark (BM) points the relative rise in entropy is shown in the following figure \ref{entropy-fig}.
\begin{figure}[h]
  \centering
  \begin{minipage}[b]{0.47\textwidth}
    \includegraphics[width=\textwidth]{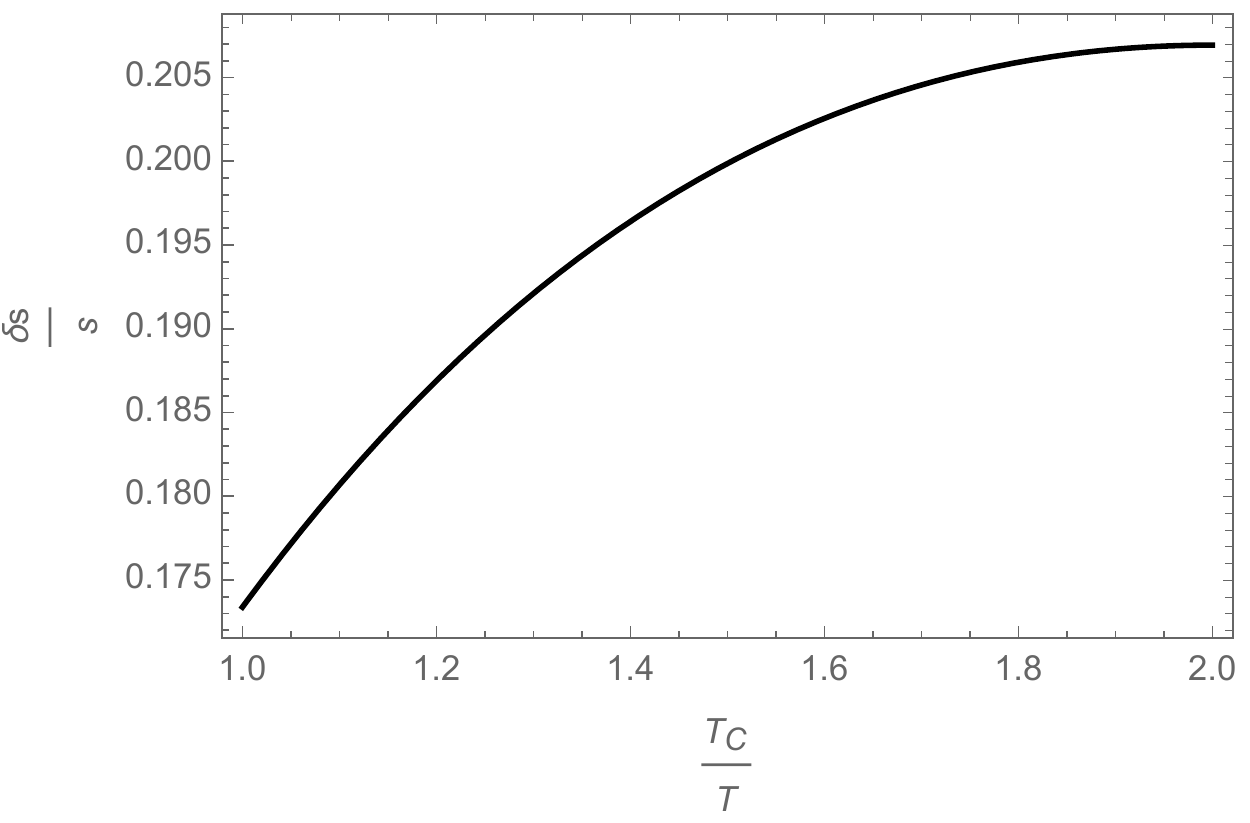}
  \end{minipage}
  \hspace*{.1cm}
  \begin{minipage}[b]{0.47\textwidth}
    \includegraphics[width=\textwidth]{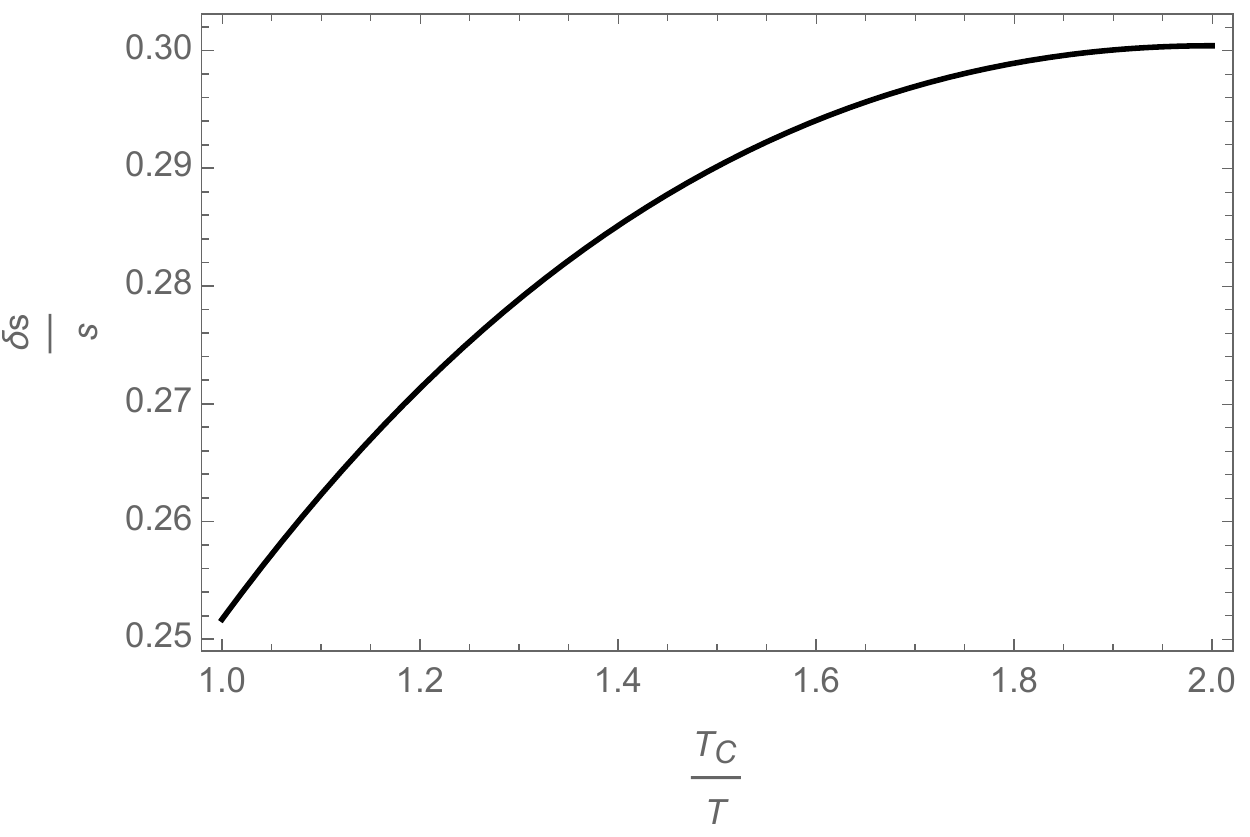}
      \end{minipage}
    \hspace*{.1cm}
    \caption{The relative increase in entropy for the two BM points are showm:\\
  \textbf{BM1}: $v_c=150.56$ GeV, $T_c=116$ GeV, $v_c/T_c=1.298$, $m_S=120$ GeV, $\lambda_m=0.77$, $\lambda_S=1$ and $\delta s/s \sim 20 \%$.\\
  \textbf{BM2}: $v_c=202.41$ GeV, $T_c=82$ GeV, $v_c/T_c=2.47$, $m_S=160$ GeV, $\lambda_m=1.35$, $\lambda_S=1$ and $\delta s/s \sim 30 \%$.\\
  As can be seen from the figure the amount of entropy influx into the plasma is dependent on the strength of the phase transition. Clearly for the BM with higher value of $v_c/T_c$ a higher entropy influx into the plasma is observed.}
 \label{entropy-fig}
 \end{figure}


\section{Conclusion}
It is shown in this paper that with the inclusion of a real singlet scalar $S$ to the SM and imposing an extra $Z_2$ symmetry, the EWPT becomes a first order phase transition. Possible benchmark points which can lead to this first order phase transition are studied and discussed in details. In the later part of the paper, the entropy release due to this first order EWPT is studied for two benchmark points numerically and shown in figure \ref{entropy-fig}. At this point two comments can be made.

Firstly, this production of entropy can highly dilute the density of any  preexisting frozen out species as well as the effective number of particles $\Delta N_{\rm eff}$. This can effect the production of stochastic gravitational waves (GW) which can be produced during EWPT. As a result the amplitude of these GWs can get diluted due to the release of entropy into the plasma during EWPT. Secondly, the stronger the phase transition is, the higher is the amount of entropy produced. The production can be even larger when a singlet scalar extension to Two Higgs Doublet Model is considered.  But these are beyond the scope of this paper and they will be studied in the subsequent papers.

\section{Acknowledgement}
The work of A.C. is funded by APEX project offered by IOP-B. J.D. acknowledges the Council of Scientific and Industrial Research (CSIR), Govt. of India for SRF fellowship grant with File No. 09/045(1511)/2017-EMR-I.


\begin{thebibliography}{99}
\bibitem{SC}
Carroll, Sean M.; Rhoades, Zachary H.; Leven, Jon (2007). Dark Matter, Dark Energy: The Dark Side of the Universe. Guidebook Part 2. Chantilly, VA: The Teaching Company. p. 59. ISBN 1-59803-350-6. OCLC 288435552.

\bibitem{Sakh}
A. D. Sakharov, JETP Letters, 1966.

\bibitem{A}
Planck Collaboration, P.A.R. Ade et al., Astron. Astrophys. \textbf{594}, A13 (2016).
[arXiv:1502.0158].

\bibitem{B}
E.W. Kolb, M.S. Turner, The early universe. Front. Phys. 69, 1–
547 (1990)

\bibitem{Higgs}
CMS Collaboration, Phys.Lett.B \textbf{716} (2012) 30-61, [arXiv: 1207.7235 [hep-ex]].

\bibitem{Chaudhuri:2017icn}
A.~Chaudhuri and A.~Dolgov,
JCAP \textbf{01} (2018), 032
doi:10.1088/1475-7516/2018/01/032
[arXiv:1711.01801 [hep-ph]].

\bibitem{shapas}
K. Kajantie, M. Laine, K. Rummukainen and M. Shaposhnikov, Phys.Rev.Lett. \textbf{77} (1996) 2887-2890, [arXiv: hep-ph/9605288 [hep-ph]].
 
 
 \bibitem{Morrissey:2012db}
D. E. Morrissey and M. J. Ramsey-Musolf,
New J. Phys. \textbf{14} (2012), 125003
doi:10.1088/1367-2630/14/12/125003
[arXiv:1206.2942 [hep-ph]].

\bibitem{z31}
M. Carena, Z. Liu and Y. Wang, JHEP \textbf{08} (2020) 107, [arXiv: 1911.10206[hep-ph]].

\bibitem{z32}
C.W. Chiang and B.Q. Lu, JHEP \textbf{07} (2020) 082, [arXiv: 1912.12634 [hep-ph]].

\bibitem{z33}
Z. Kang, P. Ko and T. Matsui, JHEP \textbf{02} (2018) 115, [arXiv: 1706.09721 [hep-ph]]

\bibitem{Chaudhuri:2021ppr}
A.~Chaudhuri and M.~Yu.~Khlopov,
Universe \textbf{7}, (2021) 8, 275
doi:10.3390/universe7080275
[arXiv: 2106.11646 [hep-ph]].
 
\bibitem{Muong-2:2021ojo}
B.~Abi \textit{et al.} [Muon g-2],
Phys. Rev. Lett. \textbf{126} (2021) no.14, 141801
doi:10.1103/PhysRevLett.126.141801
[arXiv:2104.03281 [hep-ex]].
 
\bibitem{QCD} 
T. Boeckel, S. Schettler and J. S. Bielich, 
Prog. Part. Nucl. Phys. \textbf{66} (2011) 266
[arXiv:1012.3342].

\bibitem{ACAD}
A. Chaudhuri and A. Dolgov, J.Exp.Theor.Phys. 133 (2021) 5, 552-566, 
[arXiv:2001.11219 [astro-ph.CO]].

\bibitem{ADPDIN}
A.D. Dolgov, P.D. Naselsky and I.D. Novikov, Phys. Rev. D (2000),
 [astro-ph/0009407].
 
 \bibitem{simone}
 S. Blashi and A. Mariotti, [arXiv:2203.16450 [hep-ph]].
 
 \bibitem{HN}
 P. Huet and A.E. Nelson,  Phys. Rev. \textbf{D} 53
(1996) 4578, [arXiv:hep-ph/9506477].
 
 \bibitem{BKS}
 A.I. Bochkarev, S.V. Kuzmin and M.E. Shaposhnikov, Phys. Lett. \textbf{B 244} (1990) 275.
 
 
 
 \bibitem{ACMK}
 A. Chaudhuri and M. Yu. Khlopov, Physics \textbf{3} (2021) 2, 275-289,
 [arXiv:2103.03477 [hep-ph]].
 
 
 \bibitem{ACMKSP1}
 A. Chaudhuri, M. Yu. Khlopov and S. Porey, Galaxies  \textbf{9} (2021) 2, 45,
 [arXiv: 2105.10728 [hep-ph]].
 
 
 \bibitem{ACMKSP2}
A. Chaudhuri, M. Yu. Khlopov and S. Porey, Symmetry \textbf{14} (2022) 2, 271,
[arXiv:2110.14161 [hep-ph]].

\bibitem{resum}
J. Elias-Miro, J. R. Espinosa and T. Konstandin, JHEP \textbf{08} (2014) 034, [arXiv:1406.2652 [hep-ph]].

\bibitem{18}
R. Zhou, J. Yang and L. Bian, JHEP \textbf{04} (2020) 071, [arXiv: 2001.04741 [hep-ph]].

\bibitem{19}
K. Kannike, K. Loos and M. Raidal, Phys.Rev.D \textbf{101} (2020) 3, 035001, [arXiv: 1907.13136 [hep-ph]].

\bibitem{VS}
K. Kannike, Eur.Phys.J.C \textbf{76} (2016) 6, 324, Eur.Phys.J.C \textbf{78} (2018) 5, 355 (erratum), [arXiv: 1603.02680 [hep-ph]].

\bibitem{eff}
M. Quiros, Helv.Phys.Acta \textbf{67} (1994) 451-583.

\bibitem{22}
C. Delaunay, C. Grojean and J. D. Wells, JHEP 04 (2008) 029, [arXiv: 0711.2511 [hep-ph]].

\bibitem{23}
M.E. Carrington, Phys.Rev.D \textbf{45} (1992) 2933-2944.

\bibitem{24}
K. Hashino, M. Kakizaki, S. Kanemura and T. Matsui, Phys.Rev.D \textbf{94} (2016) 1, 015005, [arXiv: 1604.02069 [hep-ph]].

\bibitem{25}
P. Bandyopadhyay and S. Jangid, [arXiv: 2111.03866 [hep-ph]].

\bibitem{26}
D. Curtin, P. Meade and H. Ramani, Eur.Phys.J.C \textbf{78} (2018) 9, 787, [arXiv: 1612.00466 [hep-ph]].
 
\bibitem{32}
A. Andreassen, W. Frost, and M. D. Schwartz, Phys. Rev. \textbf{D91} (2015), no. 1 016009, [arXiv:1408.0287[hep-ph]].

\bibitem{33}
A. Andreassen, W. Frost, and M. D. Schwartz, Phys. Rev. Lett. \textbf{113} (2014), no. 24 241801, [arXiv:1408.0292 [hep-ph]].
 
 
 
 
 \end{thebibliography}
\end{document}